\documentclass[11pt,twocolumn]{article}
\usepackage{pslatex, graphicx, amssymb, amsmath}
\usepackage{epsfig}
\usepackage{longtable}
\usepackage[font=footnotesize]{subcaption}

\def\abstract{
\typeout{Abstract}
 {\bf Abstract} 
}

\begin{document}
\title{On the influence of topological characteristics on robustness of complex networks}

\author{Dharshana Kasthurirathna\\ Centre for Complex Systems Research\\
Faculty of Engineering and IT\\
The University of Sydney\\
NSW 2006\\
Australia
        \and Mahendra Piraveenan\\ Centre for Complex Systems Research\\
Faculty of Engineering and IT\\
The University of Sydney\\
NSW 2006\\
Australia
        \and Gnanakumar Thedchanamoorthy\\ Centre for Complex Systems Research\\
Faculty of Engineering and IT\\
The University of Sydney\\
NSW 2006\\
Australia}

\maketitle
\begin{abstract}
In this paper, we explore the relationship between the topological characteristics of  a complex network and its robustness to sustained targeted attacks. Using synthesised scale-free, small-world and random networks, we look at a number of network measures, including  assortativity, modularity, average path length, clustering coefficient, rich club profiles and scale-free exponent (where applicable)  of a network, and how each of these influence the robustness of a network under targeted attacks. We use an established robustness coefficient to measure topological robustness, and consider sustained targeted attacks by order of node degree. With respect to scale-free networks, we show that assortativity, modularity  and average path length have a positive correlation with network robustness, whereas clustering coefficient has a negative correlation. We did not find any correlation between  scale-free exponent and robustness, or rich-club profiles and robustness. The robustness of small-world networks on the other hand, show substantial positive correlations  with assortativity, modularity, clustering coefficient and average path length. In comparison, the robustness of Erdos-Renyi random networks did not have any significant correlation with any of the network properties considered. A significant observation is that high clustering decreases topological robustness in scale-free networks, yet it increases topological robustness in small-world networks. Our results highlight the importance of topological characteristics in influencing network robustness, and illustrate design strategies network designers can use to increase the robustness of scale-free and small-world networks under sustained targeted attacks. 
\end{abstract}

\section{Introduction}
The study of complex networks has been one of the dominant trends in scientific research in the last decade \cite{statmech2002, alon, doro,kepes,statmech2004,PPZ, pir-plos1,pir-NHM}. Scientists from areas as diverse as physics and computer science, mathematics and biology, chemistry and social science have been interested in analysing complex networks in their respective fields and find common features between them.  A number of topological metrics have been proposed to understand the structure of a complex network: these include modularity, assortativity, information content, network diameter, and clustering coefficient among others \cite{statmech2002,doro,kepes,sole}. Meanwhile, researchers have also been interested in the functional features of networks, including their functional motifs, the routing and sharing of information over them, and how they respond to random or targeted attacks.

It has been shown that the ability of a network to maintain its integrity under node failures or attacks depends heavily on its topological structure. For example, scale-free networks are more resilient against random attacks, but more vulnerable to targeted attacks, compared to Erdos-Renyi random networks \cite{err-2000}. It can be immediately seen that quantifying such resilience (robustness) of a network is vital in a number of disciplines. For example, computer networks should be designed in such a way that they should function properly when some nodes (routers or hosts) fail, by technical faults or under attack. On the other hand, in a network of terrorist cells, we might be interested in the best strategy to attack the network so that it is disabled and disintegrated as quickly as possible.  Therefore, measuring and comparing the robustness of networks under various failure and attack scenarios is of vital importance.

Since network robustness depends on its topology, there must be quantifiable relationships between topological metrics and the robustness of a network.
In this paper, we discuss the  relationship of a particular type of robustness measure  to a number of topological parameters. The robustness measure is chosen such that it is applicable to sustained targeted attacks which follow a particular algorithm or pattern. In particular, we consider the topological features of a range of scale-free, small-world and random networks, analysing how each feature affects the robustness of each type of network.

The rest of the paper is arranged in the following manner: We will first introduce the network metrics that we will be using to quantify the topological features of a network. Then we will present the robustness metric that we will be using to quantify a network's robustness under sustained targeted attacks. In the following sections, we will present our simulation results and analyse them. Finally, we will present our conclusions.

\section{Background}

\textbf{Scale-free networks}: Scale-free networks are those networks that display similar topological features irrespective of scale \cite{statmech2002, statmech2004}. Such networks are described by power
law degree distributions, formally specified as 

\begin{equation}
\label{eq4.1} p_k = A
k^{-\gamma } U({k/k_{max}})
\end{equation}

$U$ is a step function specifying a cut off at $k = k_{max}$. The degree distribution of scale-free networks can be specified by a number of parameters, including maximum degree $k_{max}$, scale-free exponent $\gamma$, proportion of out-lier nodes $A$, and average degree $\bar{k}$. However, it can be shown that there are only two independent parameters and the others could be derived from these. 

Scale-free networks are impressively robust to random node failure and random damage \cite{doro,err-2000}. To destroy or fragment such networks randomly, one would have to remove almost all of its nodes \cite{doro}. This perhaps explains, at least partly, why scale-free architecture is commonly found in many evolved networks in nature. This also means that targeted attacks have to be designed specifically to effectively destroy such networks, and non-trivial topological analysis of the network is necessary to identify the nodes to be targeted.

Indeed, most real-world networks are scale-free networks, including technical, biological and social networks \cite{bara09, bara00, bara03, cav09, doro,mitch06a,statmech2004}. It is possible in some directed networks
that the in-degree distribution is scale-free but the out-degree distribution is not, or vice versa. For example, the in-degree distributions of some transcription networks are scale-free, while the out-degree distributions are exponential \cite{alon}. There are a number of growth models which generate scale-free networks, and prominent among them is the Barabasi-Albert model \cite{statmech2002}. 

\textbf{Small-world networks}: The small-world network model indicates that despite the large network size, the average distance between two arbitrary nodes remain relatively low. The main characteristics that are used to identify a small-world network are the relatively low average path length and the relatively high clustering coefficient\cite{statmech2004}. Indeed, many real-world networks such as the collaboration networks have been shown to possess the small-world nature. Although various models have been proposed to produce small-world networks, the most widely used one is the Watts-Strogatz model\cite{watts1998collective}. This is the model that we used to generate small-world networks in this work. Small-world networks too have been demonstrated to be robust against random node and link failures\cite{jamakovic2007influence}. 

\textbf{Erdos-Renyi random networks}: The Erdos-Renyi random model\cite{statmech2002} defines a random network where $N$ nodes are connected randomly by $M$ possible links. When generating random networks based on this model, we limited the total number of links, to make sure that the resulting network would have the predefined link to node ratio.  Even though such random networks were once used extensively to model distributed systems, researchers have since realised that most real world networks do not display degree distributions similar to random networks \cite{statmech2002}. Yet, random networks are often used as null models to compare against other networks models, and we use them for the same purpose in this work.

Now let us introduce the metrics we will use to analyse the topology of a network.

\textbf{Assortativity }: Assortativity is the tendency observed in networks where nodes mostly
connect with similar nodes. Typically, this similarity is interpreted in terms
of degrees of nodes. Assortativity has been formally defined as a correlation
function of excess degree distributions and link distribution of a network \cite{Newman2003, sole}.  The concepts of degree distribution $p_k$ and excess degree distribution $q_k$ for undirected networks are well known \cite{sole}. Given $q_k$, one can introduce the quantity $e_{j,k}$ as the joint probability distribution of the remaining degrees of the two nodes at either end of a randomly chosen link. Given these distributions,
the assortativity of an undirected network is defined as:

\begin{equation}
\label{eq1.8}\rho=\frac{1}{\sigma _q^2 }\left[ {
\sum\limits_{jk} {jk\left( {e_{j,k} -q_j q_ k} \right)} } \right]
\end{equation}

where $\sigma_q$ is the standard deviation of $q_k$. Assortativity distributions can be constructed by considering the local assortativity values of all nodes in a network \cite{EPL, EPL2}.

\textbf{Modularity}: Network modularity is the extent to which a network can be separated into
independent sub-networks. Formally\cite{alon}, modularity quantifies the fraction of links that are within the respective modules compared to all links in a network. \cite{alon} introduces an algorithm which can partition a network into $k$ modules and
measure the partitions modularity Q. The measure uses the concept that a good
partition of a network should have a lot of within-module links and a very small
number of between-module links. The modularity can be defined as:

\begin{equation}
\label{eqmod1}
Q=\sum\nolimits_{s=1}^k {\left[ {\frac{l_s }{L}-\left( {\frac{d_s }{2L}} 
\right)^2} \right]} ,
\end{equation}\\
where $k$ is the number of modules, $L$ is the number of links in the network, $l_s$ is the number of links between nodes in module $s$,  and $d_s$ is the sum of degrees of nodes in module $s$. 
To avoid getting a single module in all cases, this measure imposes $Q=0$ if all nodes are in the same module or nodes are placed randomly into modules.

\textbf{Clustering coefficient}: The {clustering coefficient} of a node characterizes the density of links in the environment closest to a node. Formally, the clustering coefficient $C$ of a node is the ratio between the total number of $y$ links connecting its neighbours and the total number of all possible links between all these $z$ nearest neighbours \cite{doro}:
\begin{equation}
\label{eqclus}
C=\frac{2y}{z\left( {z-1} \right)}
\end{equation}\\
The clustering coefficient $\overline{C}$ for a network is the average $C$ over all nodes.

\textbf{Average Path Length}: The average path length $l_G$ of a network is defined as the average length of shortest paths between all pairs of nodes in that network. For many real world networks, this average path length is much smaller than the size of the network, that is $l \ll N$. Such networks are said to be showing the small-world property \cite{lat01,new2000,wattz}. 

\begin{equation}
\label{eqavg}
l_G=\frac{1}{n (n-1)}{\sum \limits_{i,j} {d\left( {v_i, v_j} \right)}}
\end{equation}\\

Equation \ref{eqavg} gives the formal definition of the average path length of a network. Here $d (v_i, v_j)$ is the shortest path between the nodes $v_i$ and $v_j$, and $n$ is the size of the network.\\

\textbf{Rich club connectivity}: A rich-club is defined in terms of degree-based rank $r$ of nodes, and the rich-club connectivity $\varphi(r)$. The degree-based rank denotes the rank of a given node when all nodes are ordered in terms of their degrees, highest first. This is then normalised by the total number of nodes. The rich-club connectivity is defined as the ratio of actual number of links over the maximum possible number of links between nodes with rank less than $r$.  Thus, it is possible to calculate the rich-club connectivity distribution of a network, $\varphi(r)$ over $r$. Eq. \ref{eqrich} shows the formal definition of the rich-club coefficient.

\begin{equation}
\label{eqrich}
\varphi(r)=\frac{2E(r)}{r (r-1)}
\end{equation}

Here, $E(r)$ is the number of links between the $r$ nodes and $r(r-1)/2$ is the maximum number of links that these nodes could have shared.


\subsection{Topological robustness and robustness coefficient}

The ability of a network to sustain or withstand random failures or targeted attacks depends on its topological structure. For example, scale-free networks have been shown to be more resilient against random failures, but are more vulnerable to targeted centrality based attacks, in comparison to Erd\"{o}s-R\'{e}nyi random networks\cite{Albert2000}. Thus, it is important to observe the topological robustness of a network to comprehend how its topological structure would contribute to random node failures or targeted attacks.

There exists a substantial body of work which introduces and analyses structural robustness measures. Albert et al. \cite{err-2000} considered error and attack tolerance of complex networks by comparing the profiles of quantities such as the network diameter and the size of the largest component. Following their work, a multitude of metrics have been proposed to measure the topological robustness of networks as a single quantity. However, they typically  calculate averaged effects of single node removals, rather than effects of sequential removals, or are too simplistic. For example, the \textit{network efficiency} has been defined as the average of inverted shortest path lengths \cite{err-2004}, and used for quantifying the robustness of a network. Node removals are not explicitly considered in this measure. Similarly, Dekker and Colbert \cite{dstorobust} introduced two concepts of connectivity for a graph which can be used to model network robustness: the \textit{node connectivity } and \textit{link connectivity}, which are the smallest number of nodes and links respectively, whose removal results in a disconnected or single-node graph. In this work, we used the robustness coefficient introduced in Piraveenan et al. \cite{robustness} as the robustness measure as it has the advantage of providing a single numeric value to quantify the topological robustness of a network under sustained attack.


The  robustness coefficient $R$ is defined (in percentage) as \cite{robustness}:

\begin{equation}
R = \frac{200 \sum_{k=0}^{N} S_{k} - 100 S_{0}}{N^{2}}
\label{eq:robustness}
\end{equation}

In Eq. \ref{eq:robustness}, $S_{k}$ is the size of the largest component after $k$ nodes are removed. $S_{0}$ denotes the initial largest component size. $N$ is the network size. According to the above equation, for a fully connected network of any size, the robustness coefficient ($R$) would always be 100\%. While we refer the interested reader to consult \cite{robustness} for its derivation, it suffices to say here that this is essentially an Area Under Curve (AUC) measure, comparing the area under the size of largest component curve with that of a fully connected network of equivalent size. Similar Area Under the Curve (AUC) measures are used in a number of disciplines. For example, in signal detection theory, the area under  a Receiver Operating Characteristic (ROC) curve \cite{ROC} denotes the  probability that a classifier will rank a randomly chosen positive instance higher than a randomly chosen negative one, and in mechanics, the area under curve of a velocity vs time plot of a moving object denotes the distance that it has travelled \cite{kreyszig}.

Let us note that the robustness coefficient is not constant to a network, but depends on the type of attack. In order to measure the robustness coefficient under different types of attacks, the nodes that are removed are selected based on their centrality values. For example, in a betweenness based attack, the node that would be removed in each iteration would be the node with the highest betweenness centrality value, at that particular instance of the network. Likewise, the other centrality based attacks would select the node with the highest value of the respective centrality measure. In the case of a random attack, a randomly selected node would be removed in each iteration.

\section{Simulation results}

We used groups of 100 synthesised scale-free, small-world and Erdos-Renyi random networks in our simulation experiments. The scale-free networks were synthesized using a variant of the Preferential Attachment (PA) method, widely used as a model to synthesize scale-free networks \cite{statmech2002}. Each network consisted of 1000 nodes. The network size was chosen arbitrarily, while considering the computing time that would be necessary to compute the properties of a much larger network. We chose four different link-to-node ratio (LNR) values, namely $LNR=2,3,4$ and $5$. These would correspond to average degree values of  $\bar{k}=4,6,8$ and $10$. We had twenty-five networks with each LNR value in each group, and these networks were different from each other in topological characteristics such as scale-free exponent, modularity, clustering coefficient, assortativity, rich-club profiles, and average path length. 

In order to generate small-world networks, we followed the Watts-Strogatz \cite{watts1998collective} model. According to this model, initially a ring network with 1000 nodes was generated. Then for each node, $K$ number of neighbours were connected on either side of the ring. This results in a ring-lattice structure. Afterwards, based on a probability $p$, one end of each link is disconnected and connected to a random node in the ring. By varying $p$, we can vary a network between a ring-lattice structure and a random network. Values around $p=0.5$ generate small-world networks, though the values of $p$ were varied from $0.2$ to $0.8$ to achieve considerable variation in the topological parameters.  Erdos-Renyi random networks of equivalent sizes and link-to-node ratios were generated, by randomly selecting and linking pairs of nodes, to compare their robustness and other network attributes. 

When attacking the networks, we used a degree based attack. In other words, we removed the node with the highest degree form the network, in each iteration. When there are multiple nodes with the highest degree, a randomly selected node out of those is removed. In doing so, we assume that the degrees of all nodes are known prior to the attack. the choice of node degree as a node property to guide the attack is justified because this is easy to obtain or calculate and has been previously used to guide targeted attacks (e.g. in \cite{Albert2000}). All the robustness results obtained are dependent on the type of attack used. If a different type of attack was used, that would have quantitatively affected the resulting robustness values. 

We measured the robustness of each of these networks using the robustness metric described above. The topological robustness is measured as a percentage and illustrates the network's ability to withstand sustained targeted node removal. We removed the nodes in the order of node degree.

\begin{figure}[htbp]  \centering
\includegraphics[width=8.0cm]{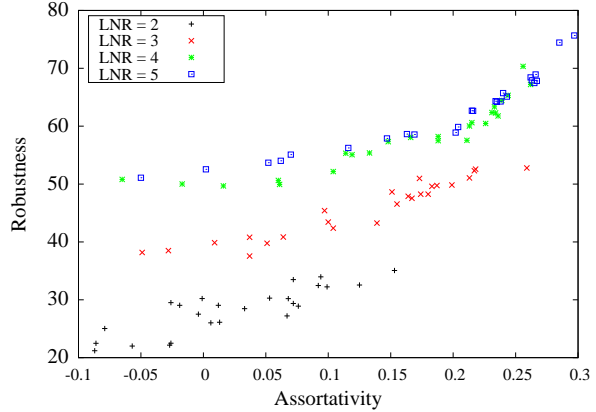} 
\caption{Scale-free network robustness against network assortativity. Four different link-to-node ratios ($LNR=2,LNR=3, LNR=4, LNR=5$) are considered.} \label{fig_assor}
\end{figure}

\begin{figure}[htbp]  \centering
\includegraphics[width=8.0cm]{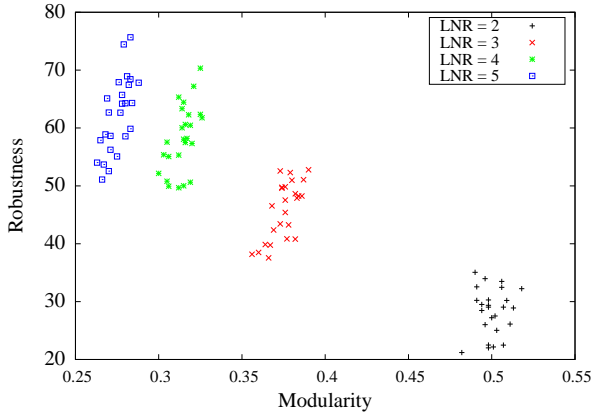} 
\caption{Scale-free network robustness against network modularity. Four different link-to-node ratios ($LNR=2,LNR=3, LNR=4, LNR=5$) are considered.}\label{fig_mod}
\end{figure}

\begin{figure}[htbp]  \centering
\includegraphics[width=8.0cm]{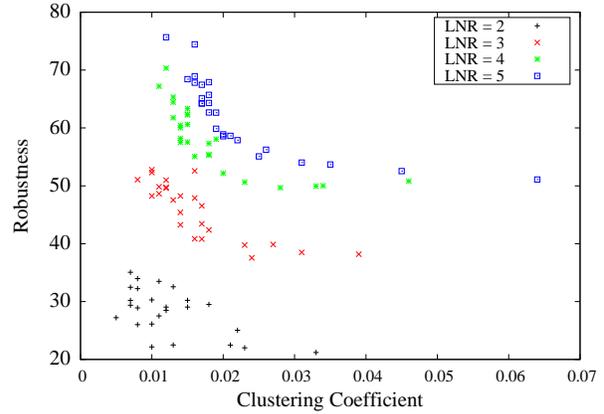} 
\caption{Scale-free network robustness against network clustering coefficient. Four different link-to-node ratios ($LNR=2,LNR=3, LNR=4, LNR=5$) are considered.}  \label{fig_cc}
\end{figure}

We now analyse how the network robustness of the scale-free networks depend on each of the topological characteristics mentioned above. Our results are illustrated in figures 1-6.  In Fig. \ref{fig_assor}, we plot the network robustness against network assortativity. The figure shows that robustness tends to increase with assortativity. This could be explained by the fact that, high assortativity means similar nodes (in terms of degrees) are connected together, including the hubs. Therefore, compared to a non-assortative scale-free network, the hubs in assortative scale-free networks have `back-ups', hence making it harder for the network to be broken apart by targeted attacks. Fig. \ref{fig_mod} shows robustness against network modularity. We find that modularity again has positive correlation with network robustness. Fig. \ref{fig_cc} shows robustness against clustering coefficient. It can be seen that from Fig. \ref{fig_cc} that these quantities are negatively correlated, with higher clustering coefficient resulting in low robustness. While this may seem counter-intuitive, it could be explained by the fact that the more clustering a network has, the less proportion of the links will be between distant parts of the network (this also explains the  negative correlation usually observed  between clustering coefficient and average path length), and as such, the network could be broken apart into components easily. While these components themselves will be quite robust to further attacks, the early disintegration means the robustness coefficient will be small for scale-free networks with high clustering.

\begin{figure}[htbp]  \centering
\includegraphics[width=8.0cm]{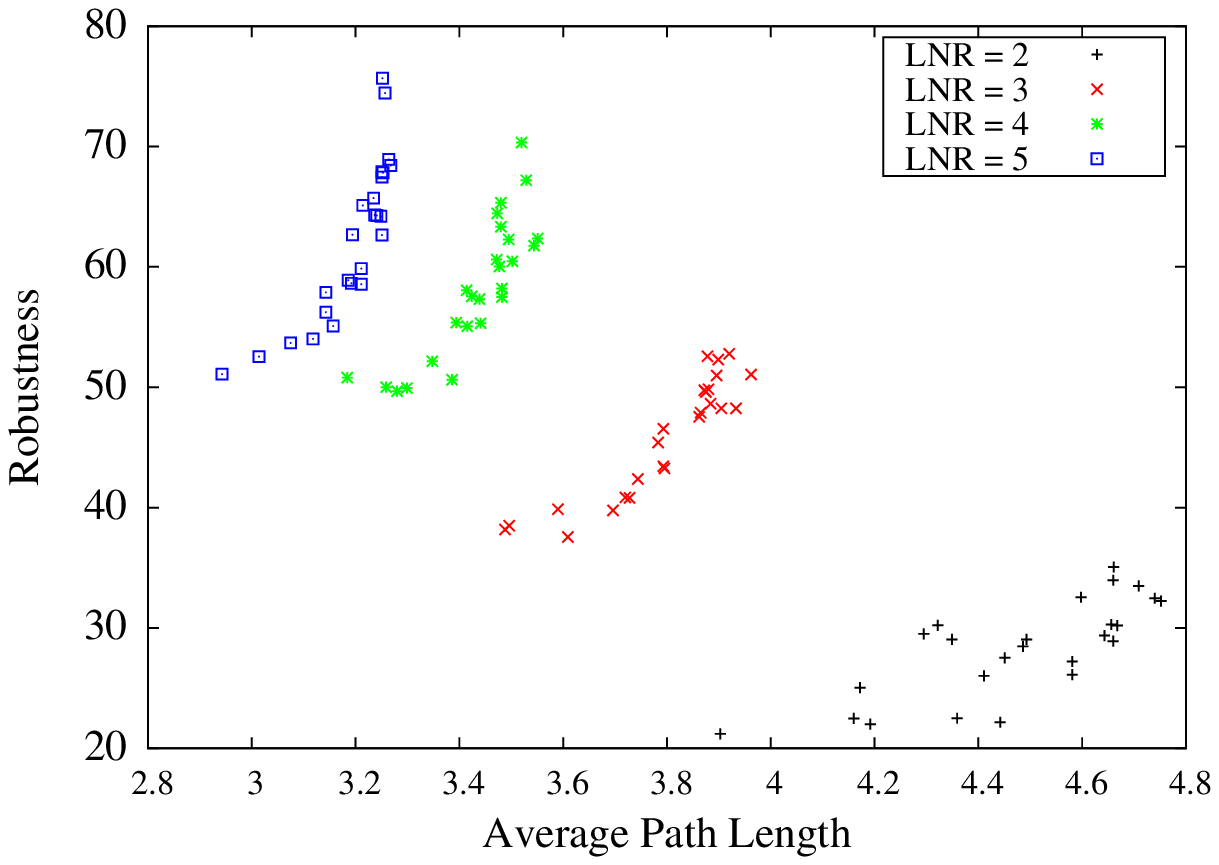} 
\caption{Scale-free network robustness against network average path length. Four different link-to-node ratios ($LNR=2,LNR=3, LNR=4, LNR=5$) are considered.} \label{fig_avg}
\end{figure}

\begin{figure}[htbp]  \centering
\includegraphics[width=8.0cm]{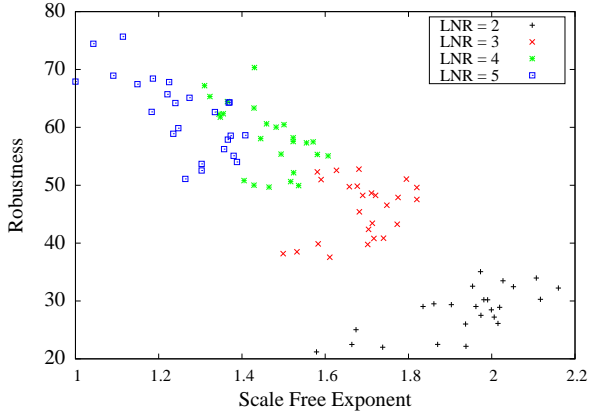} 
\caption{Scale-free network robustness against network scale-free exponent. Four different link-to-node ratios ($LNR=2,LNR=3, LNR=4, LNR=5$) are considered.} \label{fig_exp}
\end{figure}

\begin{figure*}[htbp]
\centering
	\begin{subfigure}{0.5\textwidth}
		\centering
		\includegraphics[width=6.0cm, height=4.0cm]{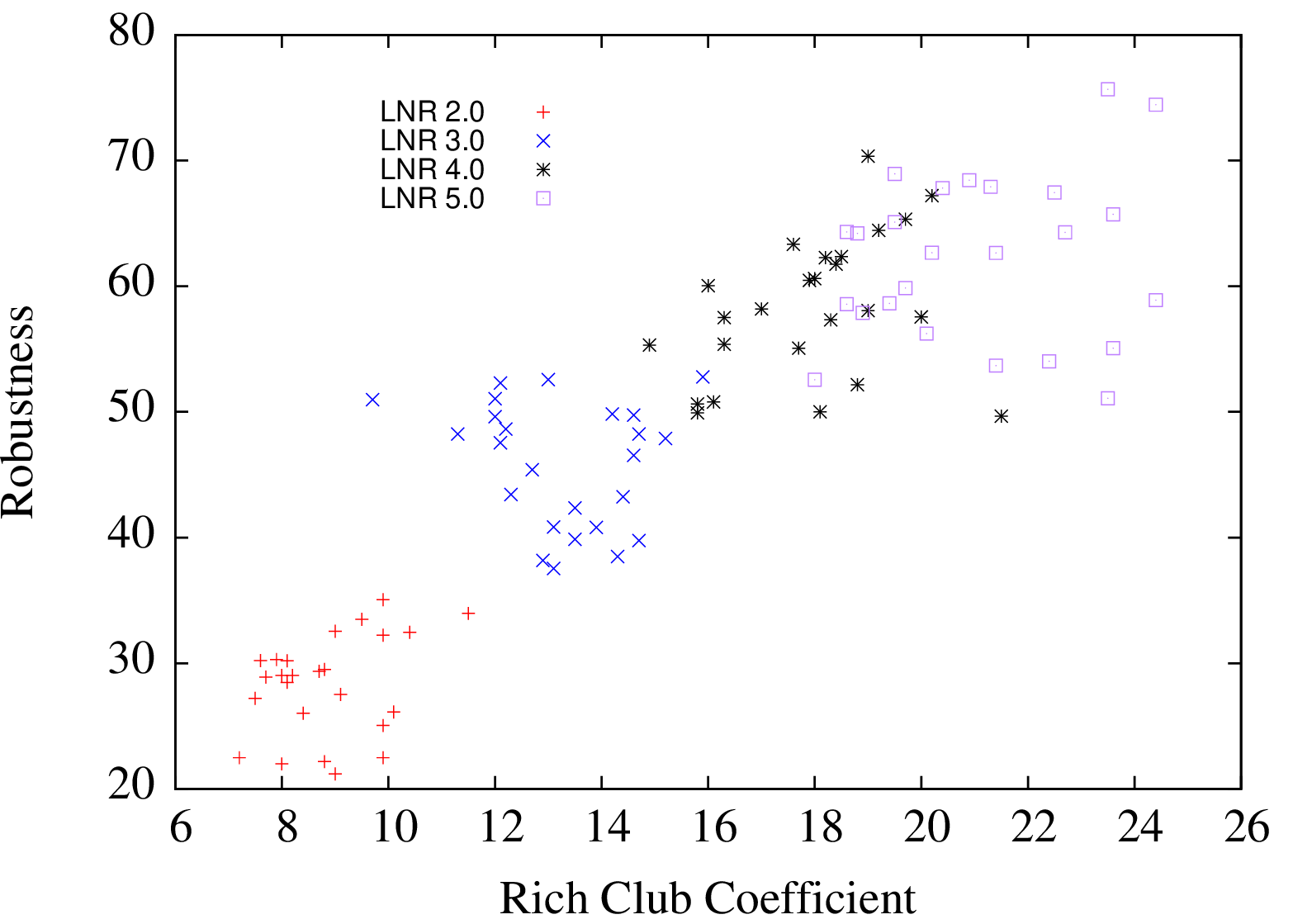}
		\caption{$Rich-Club-Threshold = 5\% $} 
	\end{subfigure}~
	\begin{subfigure}{0.5\textwidth}
		\centering
		\includegraphics[width=6.0cm, height=4.0cm]{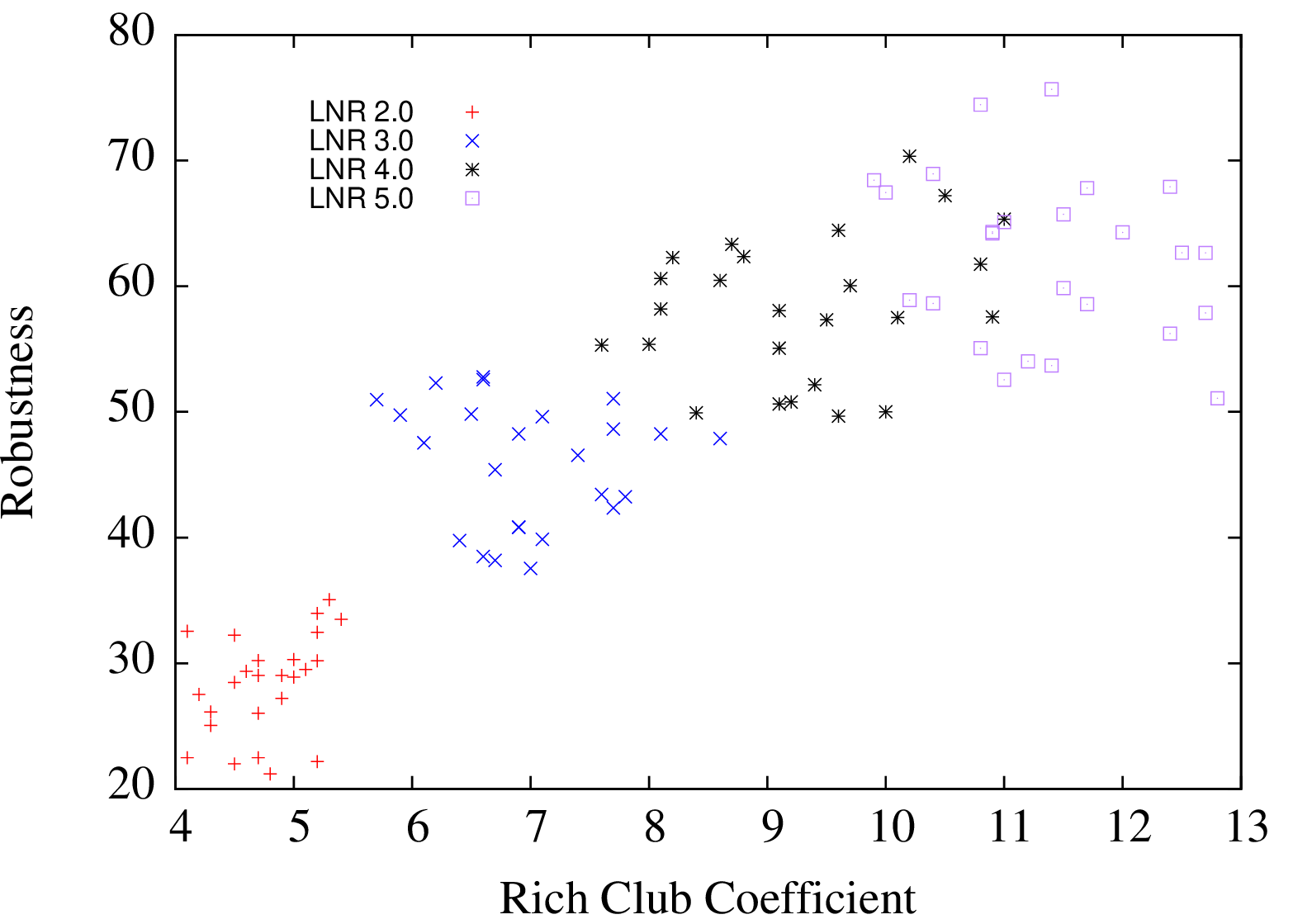}
		\caption{$Rich-Club-Threshold = 10\% $} 
	\end{subfigure}

	\vspace{15pt}
	\begin{subfigure}{0.5\textwidth}
		\centering
		\includegraphics[width=6.0cm, height=4.0cm]{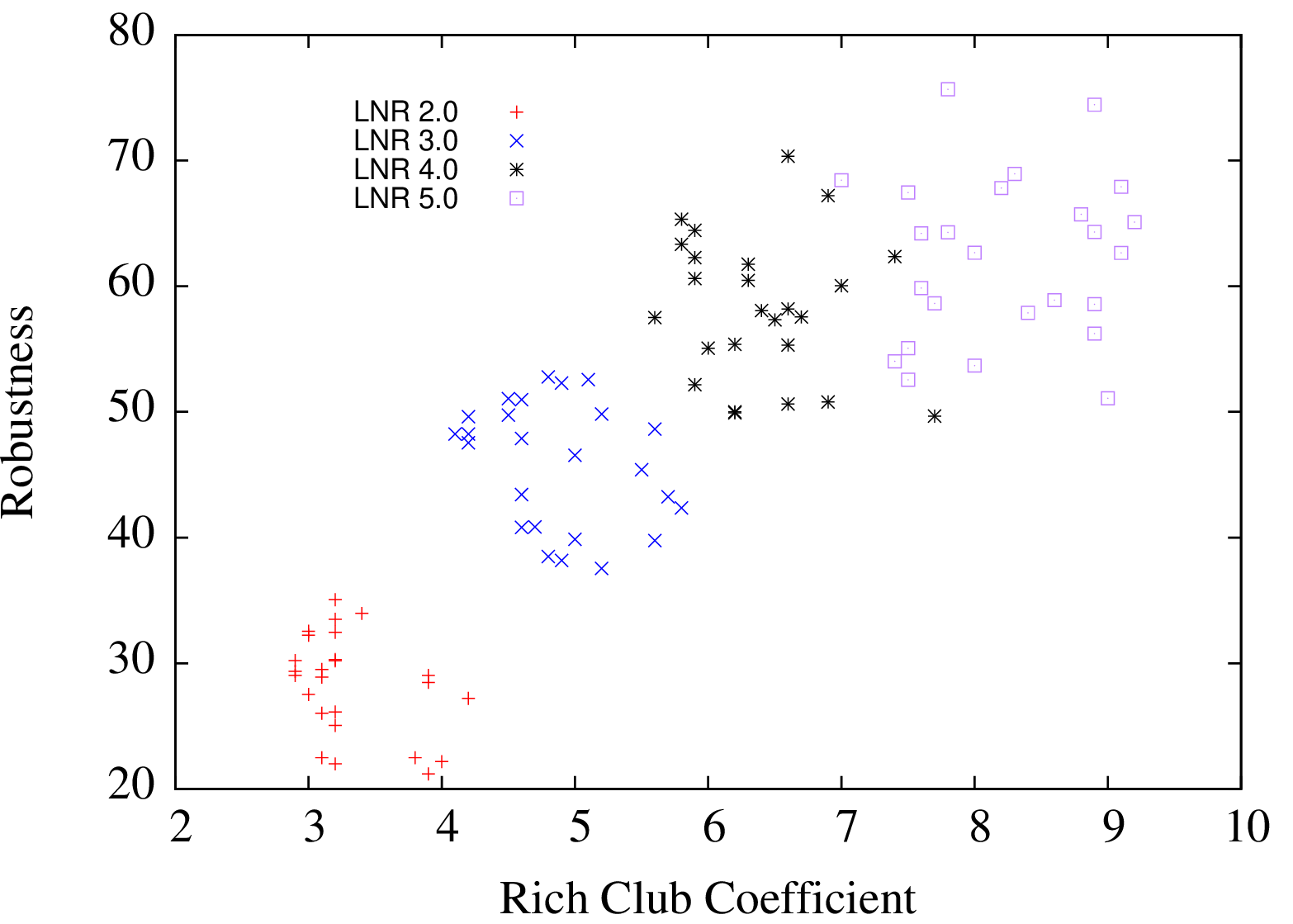}
		\caption{$Rich-Club-Threshold = 15\% $}
	\end{subfigure}~
	\begin{subfigure}{0.5\textwidth}
		\centering
		\includegraphics[width=6.0cm, height=4.0cm]{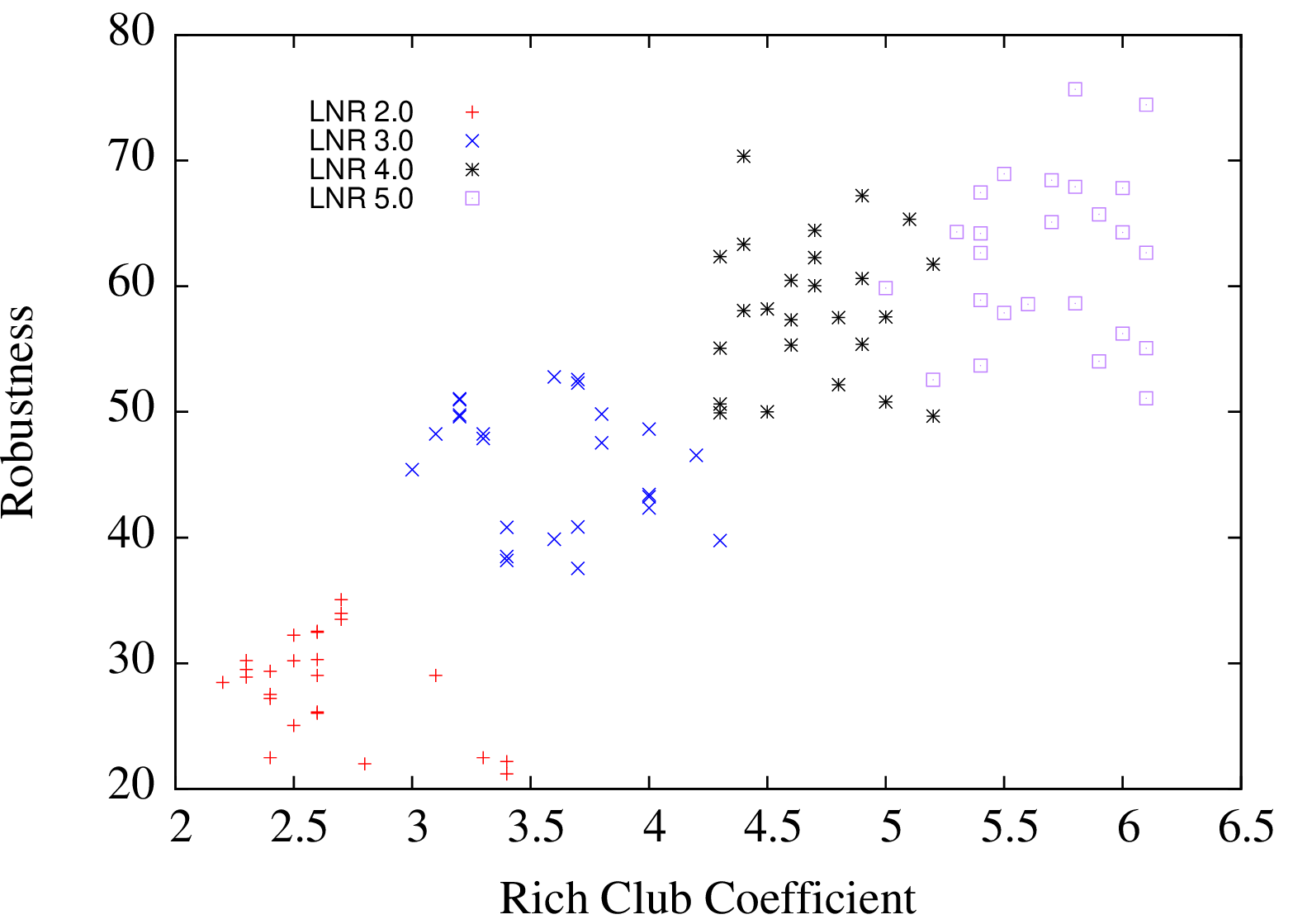}
		\caption{$Rich-Club-Threshold = 20\% $}
	\end{subfigure}
\caption{Scale-free network robustness against network rich-club coefficients. Four different link-to-node ratios ($LNR=2,LNR=3, LNR=4, LNR=5$) are considered.}  \label{fig_rich}
\end{figure*}

\begin{table}[htbp]
\scriptsize
  \raggedright
  \caption{Pearson correlation coefficients between the different network properties considered and the robustness coefficient values. Scale-free networks with four different link-to-node ratios (LNR) were considered.}
    \begin{tabular}{rrrrr} 
    \hline
          & \textbf{LNR = 2} & \multicolumn{1}{c}{\textbf{3}} & \multicolumn{1}{c}{\textbf{4}} & \multicolumn{1}{c}{\textbf{5}} \\
    \hline
    \textbf{Assortativity} & 0.86  & 0.95  & 0.90   & 0.92 \\
    \textbf{Modularity} & 0.70  & 0.70   & 0.58  & 0.70 \\
    \textbf{Clustering Coefficient} & -0.59 & -0.80  & -0.76 & -0.75 \\
    \textbf{Average Path Length} & 0.76  & 0.90   & 0.84  & 0.83 \\
    \textbf{Scale-free Exponent ($\gamma$)} & 0.71  & 0.28  & -0.55 & -0.70 \\
    \textbf{Rich Club Coefficient} &       &       &       &  \\
    \textbf{5\%} & 0.32  & -0.18 & 0.35  & 0.18 \\
    \textbf{10\%} & 0.40   & -0.11 & 0.27  & -0.25 \\
    \textbf{15\%} & -0.39 & -0.31 & -0.14 & 0.07 \\
    \textbf{20\%} & -0.43 & -0.24 & 0.04  & 0.11 \\
    \hline
    \end{tabular}%
  \label{tab:tabcorr}%
\end{table}%

Fig. \ref{fig_avg} shows network robustness against average path length. As expected (and explained above) there is positive correlation between average path length and network robustness. In Fig. \ref{fig_exp} we show network robustness against scale-free exponent. We see that the correlation pattern varies here. While it is well known that scale-free networks are more vulnerable to targeted attacks compared to random networks \cite{err-2000}, let us point out  that the scale-free exponent is \textit{not}  a measure of the \textit{scale-freeness} of a network. Therefore, the networks with high scale-free exponents are not necessarily more scale-free than those with lower scale-free exponents. Indeed, we found that in the networks we simulated, there was a  correlation between the squared error of fitting a scale-free exponent to a network, and the exponent itself, which means that the higher the scale-free exponent, the less the scale-free characteristic of a network (though the error in all cases was small enough to justify the network being identified as scale-free).

We also considered the rich-club profiles of each of the scale-free networks we generated. Since the rich-club profile is not a single quantity, we considered the rich-club coefficient of the network at four percentile values, namely 5\%, 10\%, 15\%, and 20\% (for example, a 5\% rich-club coefficient meant that the 
 sub-network consisting of the top 5\% nodes (in terms of degrees) was considered to calculate the rich club.). However, as Fig. \ref{fig_rich} shows, the rich-club phenomena does not seem to affect the robustness of networks. It may be the case that smaller rich-clubs have an effect on network robustness, and a more detailed study exploring the whole rich-club profile of each network is necessary to determine the exact effect of rich-clubs on network robustness. Such a study is beyond the scope of this exploratory paper. Table 1 summarises the results mentioned above by showing the quantitative Pearson correlation coefficients between each topological property and the robustness coefficient, for the hundred networks that we studied. In most cases, we considered values greater than 0.5, to be significant correlations. 

\begin{figure}[htbp]  \centering
\includegraphics[width=8.0cm]{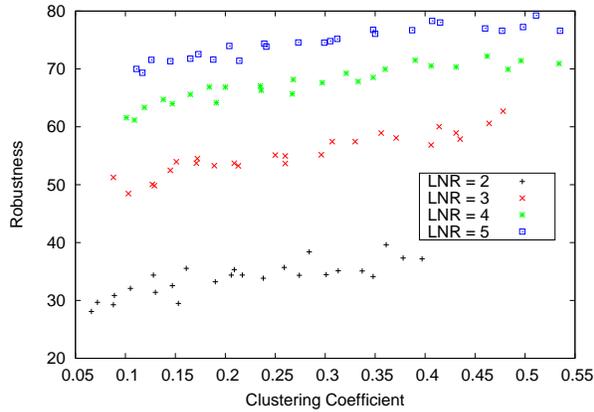} 
\caption{Small-world network robustness against network clustering coefficient. Four different link-to-node ratios ($LNR=2,LNR=3, LNR=4, LNR=5$) are considered. Contrast with Fig. 3.} \label{fig_sm_cc}
\end{figure}

\begin{figure}[htbp]  \centering
\includegraphics[width=8.0cm]{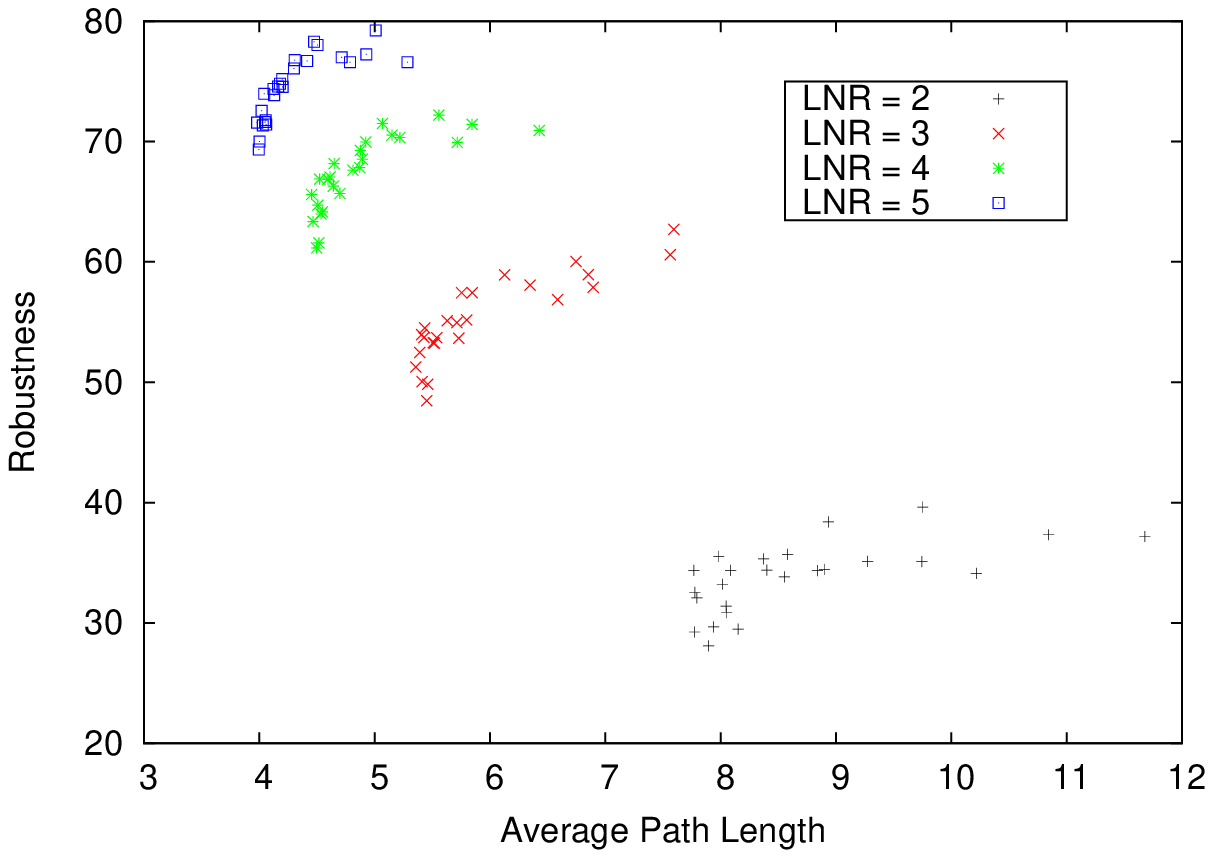} 
\caption{Small-world network robustness against network average path length. Four different link-to-node ratios ($LNR=2,LNR=3, LNR=4, LNR=5$) are considered.} \label{fig_sm_avg}
\end{figure}

\begin{figure}[htbp]  \centering
\includegraphics[width=8.0cm]{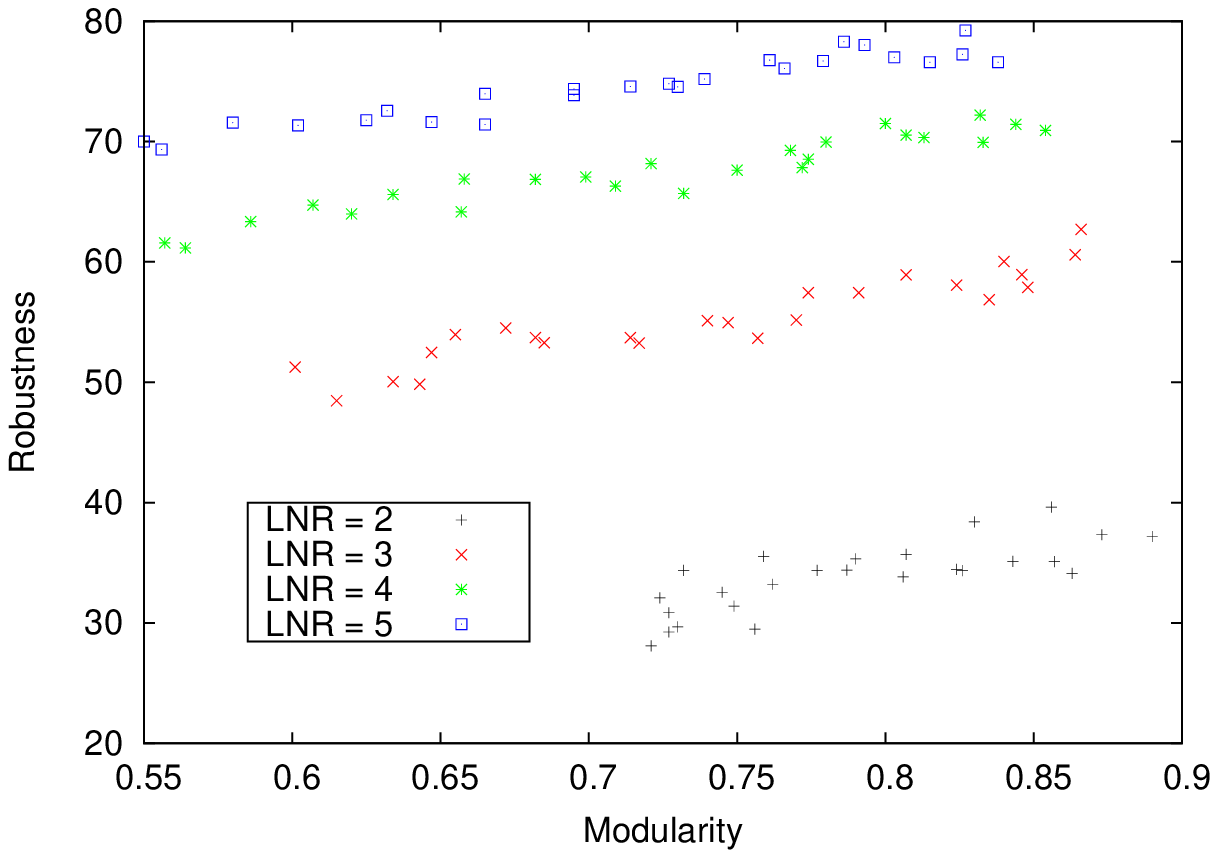} 
\caption{Small-world network robustness against network modularity. Four different link-to-node ratios ($LNR=2,LNR=3, LNR=4, LNR=5$) are considered.} \label{fig_sm_mod}
\end{figure}

Now let us turn out attention to small-world networks. Our results are summarised in Table 2. Furthermore, the figures \ref{fig_sm_cc}, \ref{fig_sm_avg} and \ref{fig_sm_mod} represent the variation of clustering coefficient, average path length and modularity against the robustness of small-world networks. These figures and table depict a positive correlation between the robustness coefficient and all topological properties considered.  What is mainly interesting here is that here the clustering coefficient too, is positively correlated with network robustness (compare Fig. 7 with Fig. 3).

\begin{table}[htbp]
\scriptsize
  \raggedright
  \caption{Pearson correlation coefficients between the different network properties considered and the robustness coefficient values. Small-world networks with four different link-to-node ratios (LNR) were considered.}
    \begin{tabular}{rrrrr}
    \hline
    \textbf{} & LNR=2 & 3     & 4     & 5 \\
    \hline 
    \textbf{Assortativity} & 0.42  & 0.34  & 0.31  & 0.43 \\
    \textbf{Modulatiry} & 0.78  & 0.93  & 0.95  & 0.95 \\
    \textbf{Clustering Coefficient} & 0.82  & 0.94  & 0.92  & 0.92 \\
    \textbf{Average Path Length} & 0.63  & 0.85  & 0.75  & 0.77 \\
    \textbf{Rich Club Coefficient} &       &       &       &  \\
    \textbf{5\%} & 0.83  & 0.93  & 0.93  & 0.93 \\
    \textbf{10\%} & 0.82  & 0.87  & 0.89  & 0.92 \\
    \textbf{15\%} & 0.81  & 0.82  & 0.84  & 0.77 \\
    \textbf{20\%} & 0.65  & 0.78  & 0.63  & 0.60 \\
    \hline 
    \end{tabular}%
  \label{tab:tblSW}%
\end{table}%

\begin{figure}[htbp]  \centering
\includegraphics[width=8.0cm]{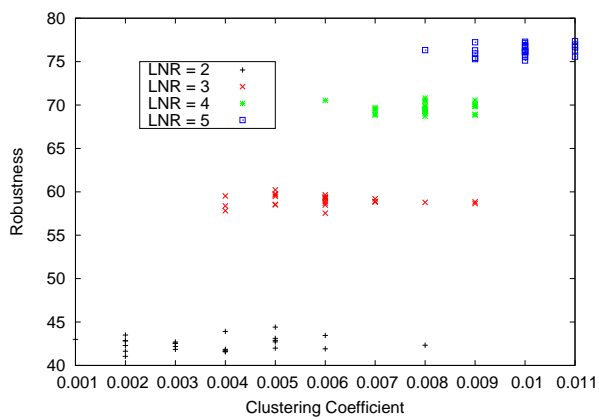} 
\caption{Random network robustness against network clustering coefficient. Four different link-to-node ratios ($LNR=2,LNR=3, LNR=4, LNR=5$) are considered.} \label{fig_ran_cc}
\end{figure}

Finally, we considered Erdos- renyi random networks. In comparison, the robustness of random networks do not show any significant correlation with the network properties considered. The figure \ref{fig_ran_cc} shows, for example, the variation of clustering coefficient with the network robustness. According to the figure, there is no significant correlation between the two parameters. Table 3 summarises our results for the correlation coefficients between the network properties considered and robustness coefficient.  Since the network is randomly linked, none of the topological properties we considered showed significant variation among various networks, and perhaps for this reason, the correlations between these properties and robustness were also mostly negligible.

\begin{table}[htbp]
\scriptsize
  \raggedright
  \caption{Pearson correlation coefficients between the different network properties considered and the robustness coefficient values. Erdos-Renyi random networks with four different link-to-node ratios (LNR) were considered.}
    \begin{tabular}{rrrrr}
    \hline 
    \textbf{} & LNR=2 & 3     & 4     & 5 \\
    \hline 
    \textbf{Assortativity} & 0.05  & 0.07  & 0.04  & 0.20 \\
    \textbf{Modularity} & 0.05  & -0.01 & 0.22  & 0.20 \\
    \textbf{Clustering Coefficient} & 0.11  & -0.09 & 0.05  & 0.22 \\
    \textbf{Average Path Length} & -0.21 & 0.05  & 0.14  & 0.12 \\
    \textbf{Rich Club Coefficient} &       &       &       &  \\
    \textbf{5\%} & -0.22 & 0.13  & 0.17  & 0.13 \\
    \textbf{10\%} & -0.11 & -0.05 & -0.01 & 0.23 \\
    \textbf{15\%} & 0.09  & -0.18 & 0.36  & -0.02 \\
    \textbf{20\%} & 0.04  & 0.24  & 0.04  & -0.02 \\
    \hline 
    \end{tabular}%
  \label{tab:tblRandom}%
\end{table}%

\section{Discussion}

Let us now discuss a few implications and limitations of our results with respect to the three types of networks that we analysed. Let us first consider scale-free networks. According to the above results, assortativity and the robustness coefficient show a clear positive correlation in scale-free networks.  Assortativity is a measure of the similarity of the nodes that are connected, in terms of the node degree \cite{new2002}. Hence, this result implies that when there are more connections among similar nodes, we can expect to see a higher topological robustness in that particular network. However, let us note that most scale-free networks we considered have positive assortativity. Those networks which were disassortative were only marginally so. Therefore, it is difficult to say whether the absolute value of assortativity has a positive correlation with network robustness.  

The next point to note is that modularity seems to substantially influence network robustness in scale-free networks. This is welcome from a network designer's point of view, since modular networks could be designed, which would also increase  the robustness of a network. Many evolved and synthesized networks in social and engineered systems are modular\cite{alon}. Furthermore, in some systems such as software networks, there is a need to design the network as a modular network \cite{robustness},\cite{Piraveenan2012}. The `functional robustness' of engineered systems also increases, it has been shown, when the networks are modular \cite{Sarkar2011}. Therefore it is significant that modular networks  have high topological robustness as well.

As mentioned above, clustering coefficient tends to have a negative correlation with average path length, when an ensemble of scale-free networks is considered \cite{pradhana}. Networks with relatively small average path length and high clustering correlation are called small-world networks \cite{statmech2002,milgram}. Our results show that the smaller the clustering coefficient is, and the larger the average path length is, the higher the robustness. This must mean that scale-free networks which also show the small-world property are relatively not robust to sustained targeted attacks. Just like scale-free networks which are not robust to targeted attacks compared to random networks \cite{err-2000, err-2004}, many natural networks which are scale-free and small-world also achieve their `small-worldness' at the cost of robustness to sustained targeted attacks.

It is already known that scale-free networks are more resilient to random node failures, compared to random networks. That is, preferential mixing is likely to increase network robustness. Preferential mixing indicates that the nodes that have higher degrees have a higher probability of attracting new links. Examples of such networks include scientific collaboration networks and social networks. Such networks generally demonstrate scale-free degree distributions \cite{statmech2002}. Note that random mixing does not necessarily imply the topology itself is random (such as an Erdos-Renyi network). Indeed, many scale-free networks, both synthesized and real world, can show near random mixing patterns \cite{Newman2003,EPL, EPL2,sole}.

With that background in mind, we may evaluate the correlation results between the scale-free exponent $\gamma$ and the robustness coefficient. When the link-to-node ratio increases, there's a deviation of the scale-free exponent from its commonly observed window of 2 to 3 \cite{statmech2002}. Moreover, the correlation between the scale-free exponent and robustness coefficient transforms from positive to negative, as the link-to-node ratio increases. This may suggest that the scale-free nature of a network could have an effect on the topological robustness of a network. However, we failed to observe a strong correlation between the scale-free exponent and the robustness coefficient. This could be partly due to the fact that we used a degree based attack to disintegrate the network, instead of emulating random node failures.

Finally, we found that in our analysis there was no correlation between rich-club tendencies and network robustness. However, we need further analysis to come to definitive conclusions here, since the rich-club phenomena cannot be measured by a single number, and is measured through a profile. Networks may show rich club phenomena at various percentile cut-offs, and we did not find that much variation among networks on the cut-offs we chose (5\%, 10\%, 15\%, 20\%). Therefore, further analysis may be necessary to establish the influence of rich-club phenomena on network robustness to targeted attacks.

In comparison, according to Table \ref{tab:tblSW} the robustness coefficient of small-world networks show high correlation with all  network parameters we considered, such as the clustering coefficient, average path length, modularity and rich-club profile. A small-world network is identified as a network, which has relatively high clustering coefficient and relatively low average path length. Also, let us note that the Watts-Strogatz algorithm that we used produces `pure' small-world networks, that do not necessarily have scale-free characters. Thus, these results affirm that the increase in average path length, which decreases the small-world behaviour, tends to increase the robustness of a network. On the other hand, when the clustering behaviour increases, that too affects the robustness in a positive manner. Even though average path length and clustering coefficient have opposing effects on the small-world nature of a network, they both seem to have a positive influence on robustness. Therefore, it is possible to `balance' the robustness of a network without losing the overall small-world nature, if we are prepared to trade off one of the small-world features (such as high clustering coefficient), for the other. Moreover, the fact that modularity and the robustness seem to show high correlation is desirable since modularity is a preferred design attribute in most designed networks, as mentioned before. Therefore, by increasing the modularity in a small-world netowrk, we can expect to preserve, if not increase its robustness. It is important to note that the correlations themselves increase when the link to node ratio increases, that is when the network gets more dense. It is also interesting to note that unlike in the case of scale-free networks, the rich club phenomena has a strong positive correlation to network robustness.

The correlation of robustness of Erdos-Renyi random networks and the network properties considered is considerably low. This may be due to the fact that the robustness coefficient values of the random networks considered did not have any considerable variation, due to the connections being random. Hence, we could argue that the scale-free and small-world network models introduce more predictability and flexibility in designing a network for its robustness, compared to random networks. This could be indeed one of the reasons why both small-world and scale-free networks are very prevalent \cite{statmech2002,alon, doro,statmech2004} in nature.

\section{Summary}

In this paper, we analysed the relationship of network robustness (under sustained targeted attacks) with a number of topological features in scale-free, small-world and random networks. Using synthesised networks, we considered topological characteristics including assortativity, modularity, clustering coefficient, average path length, rich-club profile, and  scale-free exponent. We used a particular robustness measure designed to analyse resilience under sustained targeted attacks to measure robustness. We designed our attacks based on the order of node degrees. We discussed the implications and limitations of our results.

In the case of scale-free networks, we observed substantial positive correlation between network robustness and  assortativity, modularity, and average path length in scale-free networks. Also, we observed that the clustering coefficient has a negative correlation with the robustness of scale-free networks. We did not find that  rich-club coefficients or scale-free exponents affect the robustness of a scale-free network in a significant manner. Among scale-free networks, therefore, we observed that the co-existence of small-world features (low average path length and high clustering) hinder topological robustness.

With regard to small-world networks, however, robustness coefficient showed high correlation with clustering coefficient, average path length, modularity, assortativity  and rich-club profile. However, the correlation strength with assortativity was relatively weaker compared to the rest of the parameters. We observed therefore, that high clustering increases robustness in smallworld networks yet decreases robustness in scale-free networks. In comparison, the robustness coefficient Erdos-Renyi random networks did not show any significant correlation with any of the network parameters used. We discussed the implications of our results for network design and synthesis.

\bibliography{rcr_journal}

\begin{thebibliography}{10}
\providecommand{\url}[1]{#1}
\csname url@samestyle\endcsname
\providecommand{\newblock}{\relax}
\providecommand{\bibinfo}[2]{#2}
\providecommand{\BIBentrySTDinterwordspacing}{\spaceskip=0pt\relax}
\providecommand{\BIBentryALTinterwordstretchfactor}{4}
\providecommand{\BIBentryALTinterwordspacing}{\spaceskip=\fontdimen2\font plus
\BIBentryALTinterwordstretchfactor\fontdimen3\font minus
  \fontdimen4\font\relax}
\providecommand{\BIBforeignlanguage}[2]{{%
\expandafter\ifx\csname l@#1\endcsname\relax
\typeout{** WARNING: IEEEtran.bst: No hyphenation pattern has been}%
\typeout{** loaded for the language `#1'. Using the pattern for}%
\typeout{** the default language instead.}%
\else
\language=\csname l@#1\endcsname
\fi
#2}}
\providecommand{\BIBdecl}{\relax}
\BIBdecl

\bibitem{statmech2002}
R.~Albert and A.-L. Barab{\'a}si, ``Statistical mechanics of complex
  networks,'' \emph{Reviews of Modern Physics}, vol.~74, pp. 47--97, 2002.

\bibitem{alon}
U.~Alon, \emph{Introduction to Systems Biology: Design Principles of Biological
  Circuits}.\hskip 1em plus 0.5em minus 0.4em\relax London: Chapman and Hall,
  2007.

\bibitem{doro}
S.~N. Dorogovtsev and J.~F.~F. Mendes, \emph{Evolution of Networks: From
  Biological Nets to the Internet and WWW}.\hskip 1em plus 0.5em minus
  0.4em\relax Oxford: Oxford University Press, January 2003.

\bibitem{kepes}
F.~Kepes~(Ed), \emph{Biological Networks}.\hskip 1em plus 0.5em minus
  0.4em\relax Singapore: World Scientific, 2007.

\bibitem{statmech2004}
\BIBentryALTinterwordspacing
J.~Park and M.~E.~J. Newman, ``Statistical mechanics of networks,''
  \emph{Physical Review E}, vol.~70, no.~6, pp. 066\,117+, Dec 2004. [Online].
  Available: \url{http://dx.doi.org/10.1103/PhysRevE.70.066117}
\BIBentrySTDinterwordspacing

\bibitem{PPZ}
M.~Piraveenan, M.~Prokopenko, and A.~Y. Zomaya, ``Assortative mixing in
  directed biological networks,'' \emph{IEEE/ACM Transactions on computational
  biology and bioinformatics}, vol. 9(1), pp. 66--78, 2012.

\bibitem{pir-plos1}
M.~Piraveenan, M.~Prokopenko, and L.~Hossain, ``Percolation centrality:
  Quantifying graph-theoretic impact of nodes during percolation in networks,''
  \emph{PloS one}, vol.~8, no.~1, p. e53095, 2013.

\bibitem{pir-NHM}
M.~Piraveenan, M.~Prokopenko, and A.~Zomaya, ``On congruity of nodes and
  assortative information content in complex networks,'' \emph{Networks and
  Heterogeneous Media (NHM)}, vol.~3, no. 10.3934/nhm.2012.7.441, pp. 441--461,
  2012.

\bibitem{sole}
R.~V. Sol\'e and S.~Valverde, ``Information theory of complex networks: on
  evolution and architectural constraints,'' in \emph{Complex Networks}, ser.
  Lecture Notes in Physics, E.~Ben-Naim, H.~Frauenfelder, and Z.~Toroczkai,
  Eds.\hskip 1em plus 0.5em minus 0.4em\relax Springer, 2004, vol. 650.

\bibitem{err-2000}
R.~Albert, H.~Jeong, and A.-L. Barab\'{a}si, ``Error and attack tolerance of
  complex networks,'' \emph{Nature}, vol. 406, pp. 378--382, 2000.

\bibitem{bara09}
A.-L. Barab\'{a}si, ``Scale-free networks: A decade and beyond,''
  \emph{Science}, vol. 325, no. 5939, pp. 412--413, 2009.

\bibitem{bara00}
A.-L. Barab\'{a}si, R.~Albert, and H.~Jeong, ``Scale-free characteristics of
  random networks: The topology of the world-wide web,'' \emph{Physica A}, vol.
  281, pp. 69--77, 2000.

\bibitem{bara03}
A.-L. Barab\'{a}si and E.~Bonabeau, ``Scale-free networks,'' \emph{Scientific
  American}, vol. 288, pp. 50--59, 2003.

\bibitem{cav09}
\BIBentryALTinterwordspacing
A.~Cavagna, A.~Cimarelli, I.~Giardina, G.~Parisi, R.~Santagati, F.~Stefanini,
  and M.~Viale, ``Scale-free correlations in bird flocks,'' 2009,
  arXiv:0911.4393. [Online]. Available: \url{http://arxiv.org/abs/0911.4393}
\BIBentrySTDinterwordspacing

\bibitem{mitch06a}
M.~Mitchell, ``Complex systems: Network thinking,'' \emph{Artificial
  Intelligence}, vol. 170, no.~18, pp. 1194--1212, 2006.

\bibitem{watts1998collective}
D.~J. Watts and S.~H. Strogatz, ``Collective dynamics of 'small-world'
  networks,'' \emph{Nature}, vol. 393, pp. 440--442, 1998.

\bibitem{jamakovic2007influence}
A.~Jamakovic and S.~Uhlig, ``Influence of the network structure on
  robustness,'' in \emph{Networks, 2007. ICON 2007. 15th IEEE International
  Conference on}.\hskip 1em plus 0.5em minus 0.4em\relax IEEE, 2007, pp.
  278--283.

\bibitem{Newman2003}
M.~E.~J. Newman, ``Mixing patterns in networks,'' \emph{Physical Review E},
  vol.~67, no.~2, p. 026126, 2003.

\bibitem{EPL}
M.~Piraveenan, M.~Prokopenko, and A.~Y. Zomaya, ``Local assortativeness in
  scale-free networks,'' \emph{Europhysics Letters}, vol.~84, no.~2, p. 28002,
  2008.

\bibitem{EPL2}
------, ``Local assortativeness in scale-free networks --- addendum,''
  \emph{Europhysics Letters}, vol.~89, no.~4, p. 49901, 2010.

\bibitem{lat01}
V.~Latora and M.~Marchiori, ``Efficient behavior of small-world networks,''
  \emph{Physical Review Letters}, vol.~87, no.~19, p. 198701, 2001.

\bibitem{new2000}
\BIBentryALTinterwordspacing
M.~E.~J. Newman, ``Models of the small world,'' \emph{Journal of Statistical
  Physics}, vol. 101, no.~3, pp. 819--841, November 2000. [Online]. Available:
  \url{http://dx.doi.org/10.1023/A:1026485807148}
\BIBentrySTDinterwordspacing

\bibitem{wattz}
\BIBentryALTinterwordspacing
D.~J. Watts and S.~H. Strogatz, ``Collective dynamics of small-world
  networks,'' \emph{Nature}, vol. 393, no. 6684, pp. 440--442, June 1998.
  [Online]. Available: \url{http://dx.doi.org/10.1038/30918}
\BIBentrySTDinterwordspacing

\bibitem{Albert2000}
R.~Albert, H.~Jeong, and A.-L. Barab\'{a}si, ``Error and attack tolerance of
  complex networks,'' \emph{Nature}, vol. 406, no. 6794, pp. 378--382, 2000.

\bibitem{err-2004}
P.~Crucittia, V.~Latora, M.~Marchiori, and A.~Rapisarda, ``Error and attack
  tolerance of complex networks,'' \emph{Physica A}, vol. 340, p. 388–394,
  2004.

\bibitem{dstorobust}
A.~H. Dekker and B.~D. Colbert, ``{Network robustness and graph topology},'' in
  \emph{Proceedings of the 27th Australasian conference on Computer science -
  Volume 26}, ser. ACSC '04.\hskip 1em plus 0.5em minus 0.4em\relax
  Darlinghurst, Australia, Australia: Australian Computer Society, Inc., 2004,
  pp. 359--368.

\bibitem{robustness}
\BIBentryALTinterwordspacing
M.~Piraveenan, G.~Thedchanamoorthy, S.~Uddin, and K.~S.~K. Chung, ``Quantifying
  topological robustness of networks under sustained targeted attacks,''
  \emph{Social Network Analysis and Mining}, 2013. [Online]. Available:
  \url{10.1007/s13278-013-0118-8}
\BIBentrySTDinterwordspacing

\bibitem{ROC}
J.~A. Hanley and B.~J. Mcneil, ``{The meaning and use of the area under a
  receiver operating characteristic (ROC) curve.}'' \emph{Radiology}, vol. 143,
  no.~1, pp. 29--36, Apr. 1982.

\bibitem{kreyszig}
E.~Kreyszig, \emph{Advanced Engineering Mathematics, 9th Edition}.\hskip 1em
  plus 0.5em minus 0.4em\relax {John Wiley}, December 2005.

\bibitem{new2002}
M.~E.~J. Newman, ``Assortative mixing in networks,'' \emph{Physical Review
  Letters}, vol.~89, no.~20, p. 208701, 2002.

\bibitem{Piraveenan2012}
M.~Piraveenan, S.~Uddin, and K.~S.~K. Chung, ``Measuring topological robustness
  of networks under sustained targeted attacks,'' in \emph{2012 IEEE/ACM
  International Conference on Advances in Social Networks Analysis and
  Mining}.\hskip 1em plus 0.5em minus 0.4em\relax New York: IEEE Computer
  Society, 2012.

\bibitem{Sarkar2011}
S.~Sarkar and A.~Dong, ``Characterizing modularity, hierarchy and module
  interfacing in complex design systems,'' in \emph{ASME 2011 International
  Design Engineering Technical Conferences and Computers and Information in
  Engineering Conference (IDETC/CIE2011)}, vol. 9: 23rd International
  Conference on Design Theory and Methodology; 16th Design for Manufacturing
  and the Life Cycle Conference.\hskip 1em plus 0.5em minus 0.4em\relax New
  York: ASME, 2011, pp. 375--384.

\bibitem{pradhana}
J.~T. Lizier, M.~Piraveenan, D.~Pradhana, M.~Prokopenko, and L.~S. Yaeger,
  ``Functional and structural topologies in evolved neural networks,'' in
  \emph{Advances in Artificial Life: Tenth European Conference on Artificial
  Life (ECAL '09)}, ser. LNCS/LNAI.\hskip 1em plus 0.5em minus 0.4em\relax
  Springer, 2009, vol. 5777-5778.

\bibitem{milgram}
S.~Milgram, ``The small world problem,'' \emph{Psychology Today}, vol.~1,
  p.~61, 1967.

\end{thebibliography}

\end{document}